\documentclass[preprint,prd, nofootinbib]{revtex4-1}
\usepackage{color}
\usepackage{graphicx}
\usepackage{epsfig}
\usepackage{subfig}
\usepackage{lipsum}
\usepackage{ctable}
\usepackage{hyperref}
\usepackage{physics}
\usepackage{graphicx}
\usepackage{dcolumn}
\usepackage{bm}

\begin{document}

	\title{Gravastars in a Non-minimally Coupled Gravity with Electromagnetism}

	\author{ \"{O}zcan SERT}
	\email{osert@pau.edu.tr}
	\affiliation{Department of Physics, Faculty of Arts and Sciences, Pamukkale	University,  20070   Denizli,
       T\"{u}rkiye  }
	 \author{Muzaffer ADAK}
	 \email{madak@pau.edu.tr}
	\affiliation{Department of Physics, Faculty of Arts and Sciences, Pamukkale University, 20070,  K{\i}n{\i}kl{\i},  Denizli, Turkey}
	

	\date{\today}

	\begin{abstract}

\noindent
In this paper we investigate the gravitational vacuum stars which called gravastars  in  the non-minimally  coupled models with electromagnetic and  gravitational fields. We consider two non-minimal models and find the corresponding  spherically symmetric exact solutions in the interior of the star  consisting of the dark energy condensate. Our models turn out to be Einstein-Maxwell model at the outside of the star and the solutions become the Reissner-Nordstr{\"o}m solution. The physical quantities of these models are  continuous and non-singular  in  some range of parameters and the exterior geometry continuously matches with the interior geometry  at the surface. We calculate the matter mass, the total gravitational mass, the  electric charge and redshift of the star for the two models. We notice that these quantities except redshift are dependent of a subtle free parameter, $k$, of the model. We also remark a wide redshift range from zero to infinity depending on one free parameter, $\beta$, in the second model.
	\end{abstract}
	
	\pacs{Valid PACS appear here}
	\maketitle

\section{Introduction}

Black holes are one of the most important predictions of Einstein's theory, which has been experimentally confirmed recently.
Although black holes are well known and analyzed as the exact solution of Einstein's equation, they have some properties that are not yet fully understood, such as event horizon, singularity and information paradox. One of the alternative models that can eliminate these complex properties of black holes is the gravastar (gravitational vacuum star) proposed by Mazur and Mottola \cite{Mazur2001,Mazur2004,Mazur20042}.
Gravastar is a non-singular and horizon-less ultra compact object, which has negative pressure in the interior, consisting of a dark energy condensate. Since they are ultra compact objects, it is very difficult to distinguish them from black hole observations. Although there is not sufficiently  observational evidence for the  existence of gravastars, 
it is important to investigate  them as an alternative way to better understand black holes. The increasing interest on the subject of gravastar has led researchers to  discuss and compare the  physical properties of gravastars and black holes. It has been shown that the energy flux emitted by the accretion disk surface of gravastars is smaller than the energy flux emitted by black holes and then the gravastars may have greater redshifts from black holes  \cite{Harko2009}. Thus the theoretical results might guide to observational researches in order for distinguishing them. 

The inside of the gravastar has the dark energy matter which has the equation of state $\rho=-p$, while the outside of the star is considered as a vacuum described by  Reissner-Nordstr{\"o}m  geometry. There is also an intermediate region, which consists of a thin shell containing a stiff matter satisfying the equation of state $\rho= p$. The thin shell prevents the formation of the event horizon and the phase transition surface appears at or near the region.  The surface separates the two phases of the space-time vacuum by quantum phase transition \cite{Chapline2002,Chapline2003} and then  the information paradox at black holes  is resolved by removing the event horizon. The dynamical stability of the gravastar configuration was discussed by  Visser and Witshire \cite{Visser2004}  by assuming the  infinitely small thin shell as a mathematical simplification. Then the stability was further investigated for different exterior geometries \cite{Carter2005} and dark energy star models \cite{Lobo2006}.

Moreover, even a very small electric charge relative to its mass, which may form just inside the surface of the star, prevents the nuclear matter from falling into the core  and  make the star more stable \cite{Cheng1998,Stettner,Krasinski,Sharma,mak-harko-2004}. The charged gravastar models for Einstein-Maxwell theory  were studied    in \cite{Horvat2009, Usmani2011,Bhar2014}. Then they were obtained also in the alternative gravitation models such as Born-Infeld phantom  \cite{Bilic2006} and nonlinear electrodynamics \cite{Lobo2007}. The infinitely small thin shell was replaced by a  continuous pressure profile  with an anisotropic fluid  in \cite{Cattoen2005,DeBenedic}. Meanwhile,  exact  interior and exterior solutions describing mixed relativistic stars which contain dark energy and ordinary matter were found in \cite{Yazadjiev2011}  for the model with phantom scalar field and they matched continuously  at the surface of the star.

Despite all its achievements, Einstein's theory of gravity has very important astrophysical and cosmological problems such as dark matter and dark energy. Thus, alternative theories such as $f(R)$ \cite{Chiba2003,Capozziello2006,Carroll2004,Sotiriou2010,Lobo2009} gravity,
$f(R,T)$ gravity \cite{Harko2011} etc.  were proposed to solve the challenges, where $R$ is the Ricci curvature scalar and $T$ is the trace of the stress-energy tensor. Spherically symmetric gravastar solutions for these theories were studied in \cite{Das2017,Bhar2021} for $f(R,T) $ gravity and in \cite{Shamir2018} for $f(G,T)$ gravity, where $G$ denotes Gauss-Bonnet invariant. Furthermore, the effects of the electric charge on the isotropic spherical gravastar
models were investigated in \cite{Ghosh2017} for higher dimensional Einstein-Maxwell theory and in \cite{Yousaf2019} Maxwell-$f(R,T)$ gravity by using the Mazur-Mottola conjecture. Also, a charged  gravastar model  and its stability were studied   in \cite{Ovgun2017}  by using the noncommutative geometry. Since gravastar is an ultra compact object like black hole, it has extremely dense gravitational and Maxwell fields. Thus, it is possible to consider the non-minimal $Y(R)F^2$ couplings \cite{Dereli201130111,Dereli201130112,Sert12Plus,Sert13MPLA,bamba1,bamba2,Dereli901,Baykal,Sert16regular,Dereli20113,AADS} to describe the objects. Some spherically symmetric isotropic and anisotropic compact star solutions of the theory  are investigated in \cite{Sert2017,Sert2018,Sert20182}.
In this study we give two exact non-singular gravastar solutions to the  non-minimal  $Y(R)F^2$ gravity by considering the dark energy equation of state, $p=-\rho$. This model turns out to be the Einstein-Maxwell  model at  the exterior of the gravastar and their solutions to be the Reissner-Nordstr{\"o}m solution.

\section{ A Non-minimal Model  for The Gravastars} \label{model}

In order to describe  gravastars in the non-minimal model, we consider the following  Lagrangian in differential forms  \cite{Sert2017,Sert2018}
\begin{equation}\label{action}
L  =   \frac{1}{2\kappa^2} R*1 -Y(R) F\wedge *F + 2A\wedge J + L_{mat} + \lambda_a\wedge T^a  \;
\end{equation}
where $\kappa$ is a coupling constant, $R$ is the Ricci scalar, $Y(R)$ is a function of $R$ and corresponds to a non-minimal coupling between  electromagnetism with
gravity, $A$ is the electromagnetic potential 1-form, $F=dA$ is the Maxwell 2-form, $L_{mat}$  denotes  the   matter Lagrangian  and  $J$ denotes the electromagnetic source 3-form and $\lambda_a$ denotes the Lagrange multiplier setting torsion to zero. $d$ denotes the exterior derivative, $*$ is the Hodge star operator and $\wedge$ denotes the exterior product in the exterior algebra. One can uses the Hodge map for fixing the orientation of a manifold in terms of orthonormal (Lorentzian) coframe $e^a$ by writing down the volume form $*1= e^{0123} \equiv e^0 \wedge e^1 \wedge e^2 \wedge e^3$.   

Since the search for spherically symmetric exact solutions to the gravitational field equations in the interior region of an isotropic fluid  has been a longstanding issue of theoretical physics, we assume a perfect fluid which has  the pressure $p$ and  energy density $\rho$  inside the stars, that is, 
 \begin{equation}
     \frac{\partial L_{mat}}{\partial e^a} :=  (\rho +p )u_a*u + p*e_a \ ,  \qquad \frac{\partial L_{mat}}{\partial \omega^a{}_b}:=0 \ ,
 \end{equation}
where $\omega^a{}_b$ denotes the connection 1-form and $u$ the time-like velocity 1-form, $u=u_a e^a$. Thus one can compute $\lambda_a$ from $\omega^a{}_b$-varied equation and then substitute the findings into $e^a$-varied equation of the action (\ref{action}). In conclusion we obtain the gravitational field equation 
 \begin{align}
  \label{gfe}
 - \frac{1}{2 \kappa^2}  R^{bc}
\wedge *e_{abc}  =&   Y  (F_a \wedge *F - F \wedge \iota_a *F)   +  Y_R F_{mn} F^{mn}*R_a 
\nonumber \\
&+ D [ \iota^b\;d(Y_R F_{mn} F^{mn} )]\wedge *e_{ab} + \  (\rho +p )u_a*u + p*e_a 
 \end{align}
where $R^a{}_b := d\omega^a{}_b + \omega^a{}_c \wedge \omega^c{}_b$ is the Riemann curvature 2-form, $R_b := \iota_a R^a{}_b$ denotes Ricci curvature 1-form, $\iota_a$ denotes the interior product of the exterior algebra, $D$ is the covariant exterior derivative and $Y_R := {dY}/{dR}$, $F_a := \iota_a F$, $F_{ab} := \iota_{ab}F$.
As we have shown in our previous papers \cite{Sert2017,Sert2018}, our gravitational field equation (\ref{gfe}) can be simplified considerably by the constraint  
 \begin{eqnarray}\label{cond0}
Y_R F_{mn} F^{mn} = -\frac{ k}{ \kappa^2} \ .   \end{eqnarray}
Consequently the above gravitational field equation turns into the following
 \begin{align}
\label{gfe2}
- \frac{1}{2 \kappa^2}  R^{bc}
\wedge *e_{abc}  =   Y  (F_a \wedge *F - F \wedge \iota_a *F)   -\frac{k}{\kappa^2}*R_a 
+   (\rho +p )u_a*u + p*e_a \ .
 \end{align}
We notice that as the equation (\ref{gfe})  contains higher order derivatives than two, now the new equation (\ref{gfe2}) consists of the second order differential equations. Here $k$ is an arbitrary coupling  constant between the electromagnetic and gravitational fields. It is worth to calculate the trace of this equation
 \begin{eqnarray} \label{trace}
 \frac{1-k}{\kappa^2} R = (\rho  - 3p) \;. 
 \end{eqnarray}
In this non-minimal model, while the case with $k=1$  corresponds to the radiation fluid stars with  $\rho= 3p $ \cite{Sert2017}, the other case $k\neq 1$, corresponding to $\rho \neq 3p$, has been studied in Ref. \cite{Sert2018} for certain equations of state between $\rho$ and $p$. In this study we deal with a new model for the dark energy stars defined by the equation of state  
 \begin{equation} \label{eos}
   p= - \rho  \ .
 \end{equation}
Accordingly, the Ricci scalar must be proportional to energy density via (\ref{trace}) for our model 
 \begin{equation} \label{ricciscal1}
     R= {4\kappa^2 \rho}/{(1-k)} \ .
 \end{equation}
Before ending this section, in order to close the system of equations, we write down the modified Maxwell equation by varying the action (\ref{action}) for the electromagnetic potential 1-form, $A$ 
\begin{eqnarray}
d(*Y F) = J \; .   \label{maxwell1}
\end{eqnarray}

\section{SPHERICALLY SYMMETRIC EXACT SOLUTIONS}

We study solutions with the following spherically symmetric, static line element 
\begin{eqnarray}\label{metric}
ds^2 & =& -f^2(r)dt^2  +  g^{2}(r)dr^2 + r^2d\theta^2 +r^2\sin^2\theta d \phi^2 
\end{eqnarray}
and the Maxwell 2-form representing a spherically symmetric static electric field in the radial direction   
\begin{eqnarray}
\label{electromagnetic1}
 F &= & E(r) e^1\wedge e^0 .
\end{eqnarray}
Under these ansatz our subtle condition (\ref{cond0}) takes the form 
\begin{eqnarray}\label{cond2}
\frac{dY}{dR}  =  \frac{k}{ 2\kappa^2 E^2}  \;.
\end{eqnarray}
Thus the components of the gravitational field equation (\ref{gfe}) together with the assumption of the equation state (\ref{eos}) read explicitly
 \begin{align}
\frac{1}{\kappa^2 g^2} \left(\frac{2g'}{gr}  + \frac{g^2-1}{r^2}  \right)  + \frac{k}{\kappa^2 g^2} \left(  \frac{f''}{f} -\frac{f'g'}{fg}   + \frac{2f'}{fr}  \right)  =&  YE^2   + \rho   \;, \label{gd1}\\
\frac{1}{\kappa^2 g^2} \left(
- \frac{2f'}{rf}   + \frac{g^2-1}{r^2} \right) +\frac{k}{\kappa^2 g^2} \left(\frac{f''}{f}   -\frac{f'g'}{fg}   - \frac{2g'}{rg}    \right) =& YE^2   + \rho  \;, \label{gd2}\\
\frac{1}{ \kappa^2 g^2} \left( \frac{f''}{f} - \frac{f'g'}{fg}   + \frac{f'}{rf} -\frac{g'}{rg} \right)  + \frac{k}{ \kappa^2 g^2} \left(   \frac{g'}{rg} - \frac{f'}{rf} + \frac{g^2 -1 }{r^2} \right)  =& Y E^2   -\rho  \;, \label{gd3}
 \end{align}
where prime denotes the derivative with respect to $r$.  Additionally we have checked that the exterior covariant  derivative of the gravitational field equation (\ref{gfe}) generates the conservation of the total energy-momentum tensor.
When the equations (\ref{gd1}) and (\ref{gd2}) are compared, it is seen 
\begin{eqnarray}
g(r) = {1}/{f(r)} \ .
\end{eqnarray}
Then the following two equations are left at our disposal
\begin{align}
\frac{1}{\kappa^2 } \left(  -  \frac{2ff'}{r}  + \frac{1-f^2}{r^2}  \right)  + \frac{k}{\kappa^2 } \left( ff''   + f'^2  + \frac{2ff'}{r}  \right)  =&  YE^2   + \rho   \;, \label{gd5}\\
\frac{1}{ \kappa^2 } \left( ff''  + f'^2 +  \frac{2ff'}{r}   )  + \frac{k}{ \kappa^2}(   -2\frac{ff'}{r}  + \frac{1-f^2 }{r^2} \right)  =& Y E^2   -\rho  \;. \label{gd6}
\end{align}
Here it is important to notice that the constraint equation (\ref{cond2}) is not an independent equation, because it can be derived by adding equation (\ref{gd5}) and (\ref{gd6}), then by taking  differential. Also, one can show that the trace equation (\ref{trace}) is arrived by subtracting equation (\ref{gd5}) from (\ref{gd6}).

At the end there are only three independent equations, (\ref{ricciscal1}), (\ref{gd5}) and (\ref{gd6}), in our hand to be solved, but four unknown functions $f(r), Y(R), \rho(r)$ and $E(r)$. In these cases literally two strategies are possible. One strategy is firstly to adopt a non-minimal coupling function such as $Y(R)= c_1 R + c_2 R^2$ etc, then to determine the other three unknowns. In the reverse order, one firstly assumes a metric function such as $f(r)=c_1 + c_2 r^2$ etc, and then calculates the others. In this work we follow the second approach.

\subsection{Solution-1}

A simple regular interior solution of these differential equations could be found by considering the metric function 
\begin{eqnarray}\label{g1}
f(r)  =   1 - ar^2 \;
\end{eqnarray}
where $a$ is  a real constant. Then the Ricci scalar is calculated as 
\begin{eqnarray} \label{R1}
R= 6a(4 - 5ar^2)
\end{eqnarray}
Correspondingly the other three unknown functions are computed as  
\begin{eqnarray}
E^2(r) &=& \frac{5(1+k)}{2\kappa^2C_1}a^2r^{(14k+2)/{(k+1)}}  \label{E11}
\;, \\
\rho(r)  &=&  -p(r) =  \frac{3a(1-k)}{2\kappa^2} (4 - 5ar^2)\;, \label{rho11}\\
Y(r) &=&  C_1 r^{-{12k}/{(k+1)}}  \;, \label{Y11}
\end{eqnarray}
where $C_1 $  is an integration constant which will be determined by the boundary conditions. In order for writing down the non-minimal coupling function explicitly in terms of the Ricci scalar we compute firstly the inverse function of (\ref{R1}) as $r^2= {4}/{5a} - {R}/{30a^2}$. Thus we could rewrite (\ref{Y11}) as
 \begin{eqnarray}\label{Y2}
Y(R)= C_1 \left(   \frac{4}{5a} - \frac{R}{30a^2}    \right)^{-{6k}/{(1+k)}}\;.
 \end{eqnarray} 
On the other hand, since the outside of the star is empty, it is described by the minimal Einstein-Maxwell theory and the Reissner-Nordstr{\"o}m metric which has the property $R=0$. Since we require that $Y(R)$ must be continuous, the equation (\ref{Y2}) becomes $Y(R) = C_1 ({4}/{5a})^{-{6k}/{(k+1)}}$ for $R=0$  on the boundary. For the minimal coupling, i.e. $Y(R)=1$, on the boundary this result produces
 \begin{equation} \label{C1}
  C_1=  ({4}/{5a})^{{6k}/{(k+1)}} \ .     
 \end{equation}
In summary in the interior region the non-minimal Lagrangian and the metric of the theory become, respectively,
 \begin{align}
 L_{in} =& \frac{1}{2\kappa^2} R*1 - \left(  1 - \frac{R}{24a}    \right)^{-{6k}/{(1+k)}} F\wedge *F  + 2A\wedge J + L_{mat}  + \lambda_a\wedge T^a \ , \label{theory11}\\
ds^2 _{in} =& - \left( 1- ar^2 \right)^2 dt^2 + \left( 1-ar^2 \right)^{-2} dr^2 + r^2 d\Omega^2 \ , \label{metricin}
 \end{align}
together with the electric field (\ref{E11}) to which (\ref{C1}) is inserted and the pressure (\ref{rho11}) where $d\Omega$ is the solid angle element. In the outer region the same quantities are 
 \begin{align}
 L_{out} =& \frac{1}{2\kappa^2} R*1 - F\wedge *F   + \lambda_a\wedge T^a \;, \label{theory12} \\
 ds^2_{out} =& -\left(1-\frac{2M}{r} + \frac{\kappa^2Q^2}{r^2}\right)dt^2 + \left(1-\frac{2M}{r} + \frac{\kappa^2Q^2}{r^2}\right)^{-1}dr^2 + r^2d\Omega^2 \; , \label{metricext}
\end{align}
and we have  the electric field $E(r)=Q/r^2$ and the pressure $p(r)=0$, where $M$ and $Q$ are the total mass and charge of the dark star, respectively. In the interior of  the star,  there is a specific fluid which has electromagnetic and gravitational fields with very high density. The charge  of a dark star in the volume with the radius $r$ is derived from  the modified Maxwell equation (\ref{maxwell1}) by taking the volume integral of the current density 3-form $J$
 \begin{eqnarray}\label{q1}
   q(r) = \frac{1}{4\pi}\int_V J = YEr^2\;.
 \end{eqnarray}
Here the integral is evaluated with help of the Stokes theorem. So, we calculate the total charge of the gravastar within the volume with the radius $r$ by combining the equations (\ref{E11}) and (\ref{Y11}) with the equation (\ref{q1}) 
 \begin{eqnarray}\label{qr1}
 q^2(r) = \frac{5C_1a^2(1+k)r^{{ 6(1- k) }/{(1+k)}  }}{2\kappa^2} \;.
 \end{eqnarray}
We see that $k$ must be in the range $-1 < k \leq 1$ in order for the charge to be regular at the origin. Additionally when we look at the solutions (\ref{E11}) and (\ref{rho11}), we discover that the parameter $k$ can take values in the range $-1<k<1 $.

As a result of the continuity condition of the metric, by matching the exterior and the interior metrics at the surface, $r=r_b$, we arrive at 
 \begin{eqnarray}\label{a1}
a = \frac{ 1-   \sqrt{ 1- {2M}/{r_b} + {\kappa^2Q^2}/{r_b^2} }}{r_b^2}\;.
 \end{eqnarray}
In addition, by applying  the condition that the pressure at the surface is zero, the equation (\ref{rho11}) gives one more equation among the parameters 
 \begin{eqnarray}
a = \frac{4}{5r_b^2} \label{a2}\;.
 \end{eqnarray}
Similarly, the continuity of the electric field on the boundary via the equation (\ref{q1}) with $Y=1$ yields 
 \begin{eqnarray}
Q = E(r_b)r_b^2 \label{Q12}
 \end{eqnarray}
where $Q$ denotes the total charge of the star with the radius $r_b$. If we substitute (\ref{E11}) and (\ref{C1}) into (\ref{Q12}), the the total charge could be re-expressed as
 \begin{eqnarray}\label{Q11}
  Q^2  =  \frac{ 8(1+k)r_b^2} {5\kappa^2}  \;. 
\end{eqnarray} 
We observe that the total charge increases linearly with the subtle constant $k$. We are able to calculate the total gravitational mass of the star by using (\ref{a1}), (\ref{a2}) and (\ref{Q11})
 \begin{eqnarray}\label{M1}
 M= \frac{4(5k+8)}{25}r_b \ .
 \end{eqnarray}
We also calculate the matter (or baryonic) mass of the star defined by the  integral of the energy density
\begin{eqnarray}
M_{mat} = \frac{\kappa^2}{2} \int_{0}^{r_b} \rho(r)r^2dr = \frac{8(1-k)r_b}{25} \ . \label{Mm1}
 \end{eqnarray}
  
Finally we search the gravitational redshift on the surface of the star 
 \begin{eqnarray}\label{z}
 z= (1-ar_b^2)^{-1} -1  = 4 \;.
 \end{eqnarray}
We see that all the possible $k$ values 
with different mass and charge configurations give only one redshift $z=4$. 
Experimentally high redshifts are observed in dense celestial objects such as quasars. Despite the high redshifts in general may be due to huge distance, extreme velocity or metric expansion of space, a common alternative explanation was that they were caused by extreme mass (gravitational redshifting) and electric charge. Thus, the high redshift may be due to quasars formed from ultra compact dark energy condensation at the center of galaxies, or they may be observational evidence for gravastars.

\subsection{Solution-2}

In this section we consider the following regular metric function inside the dark star
\begin{eqnarray}\label{g2}
f(r)  =  \sqrt{1 - b_1r^2 + b_2r^3}   
\end{eqnarray}
where $b_1$ and $b_2$ are arbitrary real constants. Again the equations  (\ref{ricciscal1}), (\ref{gd5}) and (\ref{gd6}) are used to determine the other unknown functions  
\begin{align}
E^2(r) =&  \frac{b_2(1+k)}{\kappa^2C_2}r^{{(11k +1)}/{(k+1})} \;,  \label{E21} \\
\kappa^2\rho(r)  =&  -\kappa^2p(r) =  (k-1)(5b_2r - 3b_1) \;, \label{rho1}\\
Y(r) =&  C_2 r^{-{10k}/{(k+1)}}   \label{Y1}
\end{align}
where $C_2 $ is an integration constant that will be determined by the boundary conditions. Now after calculating the Ricci scalar
\begin{eqnarray} \label{R}
R= 12b_1 - 20b_2r 
\end{eqnarray}
we compute the radial coordinate $r$ in terms of $R$ as $r={3b_1}/{5b_2} - {R}/{20b_2}$  and then the  non-minimal coupling function (\ref{Y1})
 \begin{eqnarray}\label{Y22}
 Y(R)= C_2 \left(\frac{3b_1}{5b_2} - \frac{R}{20b_2}\right)^{-{10k}/{(1+k)}}\;.
\end{eqnarray} 
Remembering that the exterior region of the star was described by the minimal Einstein-Maxwell theory which has the Reissner-Nordstr{\"o}m solution with the property $R=0$,  we apply the boundary condition for the non-minimal coupling function as $Y(R)=1$ for $R=0$. It determines the integration constant 
  \begin{equation} \label{C2}
 C_2=  \left( {3b_1}/{5b_2} \right)^{{10k}/{(k+1)}}
  \end{equation}
In conclusion, the Lagrangian and the metric take the forms, respectively, in the interior of the star 
 \begin{align}
 L_{in} =& \frac{1}{2\kappa^2} R*1
- \left(  1 - \frac{R}{12b_1}    \right)^{-{10 k}/{(1+k)}} F\wedge *F  + 2A\wedge J + L_{mat}  + \lambda_a\wedge T^a \ , \label{theory21} \\
ds^2 _{in} =& -\left( 1- b_1r^2 + b_2r^3 \right) dt^2 + \left( 1-b_1r^2 +b_2r^3 \right)^{-1} dr^2 + r^2 d\Omega^2 \ , \label{metricin2}
 \end{align}
and we have  the   electric field (\ref{E21}) and the pressure (\ref{rho1}). In the outer space, they are 
 \begin{align}
 L_{out} =&    
\frac{1}{2\kappa^2} R*1 - F\wedge *F   + \lambda_a\wedge T^a \ , \label{theory22} \\
ds^2_{out} =& -\left(1-\frac{2M}{r} + \frac{\kappa^2Q^2}{r^2}\right)dt^2 + \left(1-\frac{2M}{r} + \frac{\kappa^2Q^2}{r^2}\right)^{-1} dr^2 + r^2d\Omega^2 \ , \label{metricext2}
\end{align}
with the electric field $E(r)=Q/r^2$ and the pressure $p(r)=0$ where  $Q$ and  $M$  are the total charge  and mass of the dark star, respectively. By using (\ref{q1}) and (\ref{E21}), we calculate the electric charge of the star inside the sphere with radius $r$ 
 \begin{eqnarray}\label{qr2}
 q^2(r) =  \frac{C_2 b_2 (1+k)}{\kappa^2}  r^{-{5(k-1)}/{(1+k)} }\;.
 \end{eqnarray}

Continuities of the metric functions and the pressure on the boundary surface at $r=r_b$ generate 
 \begin{align}
 1-b_1r^2 +b_2 r^3 =& 1- \frac{2M}{r_b} + \frac{\kappa^2Q^2}{r_b^2} \;, \label{a22}\\
 b_2 =& \frac{3b_1}{5r_b} \;. \label{b2}
 \end{align}
By inserting (\ref{C2}) and (\ref{b2}) into (\ref{qr2}) on the surface we can calculate the total charge of the star $Q$  
 \begin{eqnarray}\label{Q22}
 Q^2  =  \frac{3\beta (1+k)r_b^2}{5\kappa^2}
 \end{eqnarray} 
where we defined a new dimensionless constant $\beta$ instead of the physically dimensional constant $b_1$ as $\beta  = b_1r_b^2$. We remark that the increasing $k$ and $\beta $ values increase the total charge. Besides we calculate the total gravitational mass which is the mass appeared in the metric (\ref{metricext2}) and the matter mass defined by the volume integral of the energy density (\ref{rho1}) over the star as follows
\begin{eqnarray}
 M&=& \frac{(3k+5)\beta }{10} r_b \;, \label{Mm2} \\
 M_{mat} &=& \frac{(1-k) \beta }{8}r_b \;. \label{matmass2}
 \end{eqnarray}

Besides, the gravitational redshift for the surface of the star is computed as 
 \begin{eqnarray}\label{z2}
 z= (1-b_1r_b^2 + b_2 r_b^3)^{-1} -1  = \frac{1}{1-\frac{2}{5} \beta } -1 \; .
\end{eqnarray}
Compared with our previous model which has only one free parameter $k$, in this model we have a new arbitrary parameter $\beta$ along with $k$. As usual the observational extreme values will guide us to determine the limits of $\beta$.
According to the literature \cite{F_Wang2021, B_W_Lyke2020, E_Momjian2021}  more than 750,000 quasars have been found as for today. All observed quasar spectra have redshifts between 0.056 and 7.642.  Accordingly we obtain the interval for $\beta$ as $0< \beta <2.47$.
In addition, a much larger redshift can be obtained in the range  $2.47 \leq \beta <2.5 $. Future observation results in this direction may help determine the parameter $\beta$ for each dark quasar.

\section{Conclusion}
 
We have constructed two  exact, new gravastar  solutions  to the non-minimally coupled $Y(R)F^2$ theory.  The interior of the gravastar consists of  the  dark energy condensate which has the equation of state $p=-\rho$ and the exterior is vacuum describing  by the Reissner-Nordstr{\"o}m geometry. Then the  thin  shell   proposed by Mazur and Muttola \cite{Mazur20042} is replaced by  an infinitely thin surface,  which continuously connects these two regions. Thus, in this study, all the physical quantities are continuous and the exterior geometry continuously matches with the interior geometry on the surface.
We obtain the total gravitational mass, matter mass, total electric charge and gravitational surface redshift in terms of the  free parameters of the model. 
In the first solution we have only one free parameter $k$ and in the second solution two parameters $k$ and $\beta$. We determined the ranges of these parameters from regularity of the physical quantities.
The second solution gives a wide redshift range from zero to infinity depending on the $\beta$ parameter, while the first solution gives a single redshift value. In the literature, although there is an upper limit of the surface redshift for uncharged general relativistic objects as $z<2$ \cite{Buchdahl}, the surface redshift can reach much higher values such as $z=5$ \cite{Bohmer-Harko2006} and $z=5.211$ \cite{Ivanov2002,Guven1999} for anisotropic stars. 
Furthermore, this limit can be exceeded in the charged case and it is proportional to  the charge parameter $Q$ \cite{Mak1}. Moreover, by considering alternative gravity models such as Kaluza-Klein theory this limit can reach much higher values \cite{Zhang2006} which is very close to our result $z=4$. The authors of a very recent paper inform the discovery of a luminous quasar at $z = 7.642$, J0313-806 and the most distant quasar observed so far \cite{F_Wang2021}.
Besides, the surface redshift can be  indefinitely large  for quasiblack holes \cite{Arbanil2014,Lemos2004,Bronnikov2014}.
Therefore, these high redshift values in our model may be due to  quasars, quasi-black holes or gravastars formed from ultra compact dark energy condensation at the center of galaxies.




\end{document}